\documentclass[9pt, twocolumn, a4paper]{scrarticle}

\usepackage[T1]{fontenc}
\usepackage[utf8]{inputenc}
\usepackage[separate-uncertainty=true, per-mode=symbol]{siunitx}
\usepackage{graphicx}
\usepackage[hidelinks]{hyperref}
\usepackage{cite}

\begin{document}
\title{Laser Scanning Microscopy for Tomographic Imaging of\\ Roughness and Point Absorbers in Optical Surfaces}
\author{Leif Albers, Malte Hagemann, and Roman Schnabel\\
	\\
	Institut für Quantenphysik und Zentrum für Optische Quantentechnologien\\
	Universit\"at Hamburg, Luruper Chaussee 149, 22761 Hamburg, Germany\\
	\\
	Leif.Albers@physik.uni-hamburg.de\\
}
\date{19.12.2023}
\maketitle

\section{Abstract}
High-precision laser interferometric instruments require optical surfaces with a close to perfect contour, as well as low scattering and absorption. Especially point absorbers are problematic because they heat up at high optical intensities and locally deform the otherwise flat surface, resulting in correlations between absorption and contour. Here, we present a laser scanning microscopy approach for the reconstruction of the two complementary images of an optical surface. The 'phase image' is related to the surface profile including roughness. The 'loss image' localizes point absorbers. Our experiment achieves a sensitivity of up to \SI{3.1(1.4)}{\femto\meter\per\sqrt{\hertz}} and a \SI{5.29(6)}{\micro\meter} lateral resolution. The two images show correlations for some features proving the particular strength of our tomographic approach, which should help further improving optical surfaces or to understand  dynamic processes of surface physics.

\section{Introduction}
Precision optics play a significant role in scientific research and technological advancements, ranging from laser gyroscopes \cite{Schreiber2011HowTD, skeldon2001qualification} to the detection of gravitational waves \cite{Driggers2015AdvancedL, beauville2004virgo, sato1999loss}. The quality of optical components critically influences the performance of these systems. In particular, the surface roughness of substrates and coatings of laser mirrors is a limiting factor, as it produces undesired scattering \cite{elson1980light}. Another limitation is absorption, especially in the form of localized absorbing features, introducing thermal distortions at high light powers \cite{jia2021point, brooks2021point}. Profiling and characterizing these attributes with high sensitivity and resolution and finding their correlations are essential for optimizing optics and instruments.

The most commonly used and commercially available techniques for imaging surface roughness are atomic force microscopy and variations of interference microscopy \cite{chkhalo2014roughness, gong2018surface}. While atomic force microscopy allows for precise surface topography, offering up to atomic resolution, it is typically limited to a small field of view (FOV) and can be quite time-consuming \cite{chkhalo2014roughness, gong2018surface, eaton2010atomic}. Interference microscopy methods allow for fast profiling of larger FOVs, however, since they involve a two-dimensional imaging sensor of a fixed pixel density \cite{de2011coherence, de2015principles, wyant2002white}, they exchange an increase in FOV area for a decrease in lateral resolution. Additionally, both methods typically specialize in measuring one of the two observables of the surface's phase space, i.\,e. they are unable to detect both the surface profile and optically absorbing features. They are thus unable to detect correlations between them.

We present a fast laser scanning microscopy technique combined with computational image reconstruction, offering up to diffraction limited lateral resolution and interferometric sensitivity independent of the FOV size. It maps optical components with respect to reflection phase and optical loss, by scanning a focused probe laser over their surface in an efficient spiral pattern, decoupling resolution from FOV size and allowing for fast measurements.
Images of the substrate surface of a high-quality laser mirror taken in quick succession show correlations between loss and phase. In principle, both images can also be measured simultaneously and dynamic correlation processes can be tracked. If the complementary information is measured at the same time, the quantum noise of the individual images only increases by a factor of two in the noise power.

\section{Experiment}
\begin{figure}[htb]
	\centering
	\includegraphics[width=7.5cm]{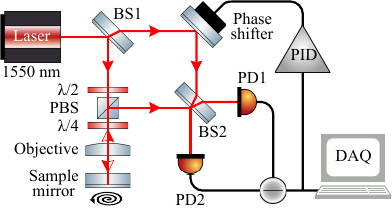}
	\caption{\textbf{Schematic of the experiment}. BS1 and BS2 were balanced beam splitters forming a Mach-Zehnder interferometer. An objective lens focused the measurement beam through the sample mirror onto its reflection coating. The lens was corrected for the substrate thickness. The reflected beam was isolated using a combination of a polarizing beam splitter (PBS) and a quarter-wave plate. The interferometer was servo-controlled on half fringe and read out by recording the difference voltage of two photo diodes (PD) with a data acquisition card (DAQ).}
	\label{fig:setup}
\end{figure}

Our setup was based on a Mach-Zehnder interferometer, as illustrated in figure\,\ref{fig:setup}. We used a \SI{0.5}{\milli\watt} continuous-wave monochromatic source beam in the TEM$_{00}$ mode with center wavelength of \SI{1550}{\nano\meter}, supplied by a NKT Basik Laser module. It was split equally at a balanced beam splitter (BS1) into a signal path (reflection) and a reference path (transmission). Using a microscope objective, the signal path beam was strongly focused under normal incidence onto the substrate surface of a high-reflectivity mirror from its anti-reflection-coated side. For the beam's intensity profile, we determined a full Gaussian width of \SI{5.29(0.06)}{\micro\meter}, defining our setup's lateral resolution. The technique's resolution is fundamentally constrained only by diffraction.

The sample mirror consisted of a planar fused-silica substrate of \SI{25.4}{\milli\meter} diameter and \SI{6.35}{\milli\meter} thickness with a specified root-mean-square (rms) roughness of $<\SI{0.8}{\nano\meter}$. It had a high-reflection coating with a specified reflectivity of \SI{99.95}{\percent} for \SI{1550}{\nano\meter} applied by 3photon via ion beam sputtering. The back was anti-reflection coated. The technique is not constrained to highly reflective optics.  Our objective was designed by ASE OPTICS to illuminate a sample through a \SI{6.35}{\milli\meter} thick fused silica coverplate. Utilizing this, we illuminated the sample mirror from the back to measure the coating profile at the interface between substrate and high reflection coating. Therefore, the measurement was free from external impurities, such as dust particles adhering to the mirror, and provided information of the substrate's effective roughness. The setup is not limited to this configuration, if a different objective design is used.

The sample mirror was mounted to a combination of two motorized high-precision stages. A rotation stage spun the sample mirror at a frequency of \SI{5}{\hertz}, while a linear stage moved both radially with \SI{10}{\micro\meter\per\second}. Combined, they drove the sample mirror along a spiral scanning trajectory, mapping circular and annular FOVs. To prevent gaps in the scan, the radial velocity needs to be smaller than one focus diameter per revolution. Simultaneously, the sampling rate must be sufficiently high to ensure that the sample spacing is always smaller than one beam diameter. The scanning duration then scales linearly with the radius of the FOV at fixed radial velocity and rotation frequency.

Because the stacked stages elevated the sample mirror well above beam height, a stable periscope was built to enable normal incidence. Upon reflection, the focused probe beam acquired a microscopic reflection phase, depending on the height profile of the sample mirror's coating. The beam carrying this phase signal is isolated using a combination of a polarizing beam splitter and a quarter-wave plate.  Two differential photodiodes measured the interference created by overlapping signal and reference path at a second beam splitter (BS2), detecting the microscopic phase shift. While moving the stages, we measured an average fringe visibility of \SI{96}{\percent}. The differential voltage was recorded by a data acquisition card with \SI{3.571}{\mega Sa\per\second} and 16-bit resolution. Additionally, it was filtered by the PID and fed back to a mirror mounted to a piezo actuator (phase shifter), locking the interferometer to midfringe for optimal phase sensitivity\cite{dowling2014quantum}.  The phase shifter consisted of a mirror mounted to a piezo actuator. We ramped the phase shifter to determine the slope at midfringe for calibration of photocurrent to reflection phase.

\begin{figure}[t]
	\centering
	\includegraphics[width=\linewidth]{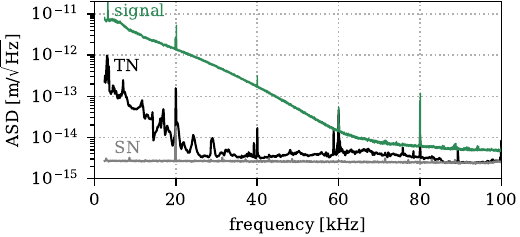}
	\caption{\textbf{Measurement data.} Amplitude spectral density (ASD) of the interferometer total noise (TN) measured while the sample position was stationary in comparison to shot noise (SN) and signal ASD measured while scanning the surface. All traces were averaged 1000 times. TN and SN have a resolution bandwidth of \SI{53}{\hertz} and the signal \SI{5}{\hertz}. Below \SI{60}{\kilo\hertz} the average signal-to-noise ratio calculates to \SI{32}{dB}}
	\label{fig:sensitivity}
\end{figure}

In the signal's spectral range, the interferometer reached a sensitivity of up to \SI{3.1(1.4)}{\femto\meter\per\sqrt{\hertz}}, which was close to the shot noise limit of \SI{2.5(1.2)}{\femto\meter\per\sqrt{\hertz}}. This can be seen in figure\,\ref{fig:sensitivity}. Although the rotation frequency of the scan was \SI{5}{\hertz}, the signal's spectrum ranged to above \SI{60}{\kilo\hertz}, as high frequency components corresponded to a steep slope in the measured phase profile. The steepest detectable slope was limited by convolution with the beam profile. Due to the spiral scanning, the probe beam had a large velocity relative to the sample mirror. Combined with the strongly focused spatial profile, this resulted in a very narrow temporal beam profile. Consequently, the beam had a broad spectral profile, enveloping the reflection phase spectrum. Since the relative beam velocity depends on the radius and continuously changes during the scan, the final spectrum is not Gaussian but the integral over a family of Gaussian profiles of increasing width.

The measured phase depends on the atomic surface profile of the coating and volumetric effects in the innermost coating layers, like modulations of the refractive index. It is specific to the wavelength used (here \SI{1550}{\nano\meter}) and coating properties, directly revealing the mirror's influence on reflected light and therefore carries great significance for practical applications at the same wavelength.

Alternatively to the reflection phase, the almost identical setup was used to measure the spatially resolved optical loss of the laser mode. With the differential photodiodes replaced by a single photodiode in the signal arm, we recorded an optical loss map of the same FOV, resolution, and topology as the phase map, providing information of surface scattering and absorption. By splitting the reflected light beam for two complementary observations, future implementations could measure two observables simultaneously.

\section{Data processing}
In the interest of data management and computation time, we processed all recorded data revolution-vise. The dominant source of noise were vibrations due to mechanical imperfections in the motorized stages. To suppress these contributions, a spectral filter was applied, removing all frequency components below a threshold of \SI{2.5}{\kilo\hertz} and \SI{0.5}{\kilo\hertz} for the phase and loss measurements, respectively. This proved to be very effective in removing noise and minimizing artifacts which inevitably appear around steep signal slopes and at the edges of data sets. Additionally, this imposed a lower limit to the detectable spatial frequencies of about \SI{18}{\per\milli\meter}. Reducing mechanical noise and consequently lowering the filter threshold would improve this limit.

From the measured reflection phase, we derived an effective surface level relative to the profile mean, corresponding to the signal beam's one-way change in the propagation distance within the substrate. The reconstructed map thus shows a height profile and allows for calculation of the effective rms roughness. We use the term effective roughness to emphasize the contribution of volumetric effects and the convolution with the beam profile.

Optical loss per beam area was approximated by normalization to the mean value of the entire scan area. To correct for detection noise, the rms error was subtracted in advance. Calibrating the photodiode would enable measuring the absolute optical loss.

As the recorded one-dimensional dataset contained no explicit spatial dependency, the position of each sample was calculated from an ideal spiral trajectory and binned into image pixels. Due to choosing our sampling rate relatively high, each individual pixel covered multiple samples. Therefore, we took the arithmetic mean of these samples. We also calculated the statistical deviation (rms error) per pixel, yielding an uncertainty map. The pixel density is independent of the lateral resolution and is only limited by the sampling rate.

\section{Results}
\subsection{Interferometric phase map of substrate to coating interface}
\begin{figure}[htb]
	\centering
	\includegraphics[width=\linewidth]{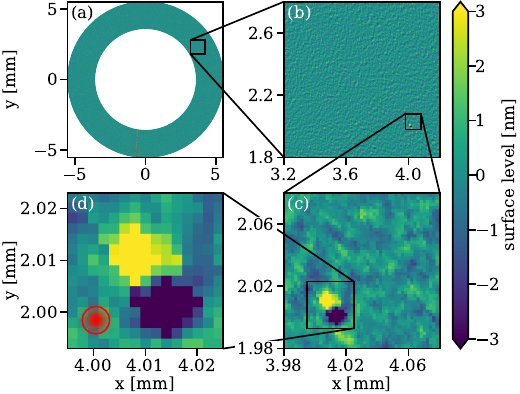}
	\caption{\textbf{Reconstructed map of measured reflection phase.} Reflection phase was measured per beam area $\approx\SI{22}{\micro\meter\squared}$ and is expressed in terms of surface level with a conversion factor of \SI{11.71}{\milli\radian\per\nano\meter}. (a) Overview of a full annular scan of \SI{2}{\milli\meter} thickness. (b) Section of (a), enlarged by a factor of 10, showing the effective roughness profile of the sample mirror. (c) Section of (b), enlarged by a factor of 10, showing a microscopic phase feature. (d) Section of (c), enlarged by a factor of 3, comparing the feature to the focus diameter, indicated by a red intensity Gaussian and a circle corresponding to its waist. Lateral resolution is \SI{5.29}{\micro\meter}.}
	\label{fig:effective_roughness_map}
\end{figure}

Fig.\,\ref{fig:effective_roughness_map} shows a reconstructed reflection phase profile map scanned from \SI{3.5}{\milli\meter} to \SI{5.5}{\milli\meter} radius, at the time limited only by alignment of the linear stage. It covers an area of about \SI{56}{\milli\meter\squared}, corresponding to more than \SI{11}{\percent} of the sample mirror's surface area. The measurement took only \SI{200}{\second} and is repeatable even after realigning the setup days apart.

Fig.\,\ref{fig:effective_roughness_map}\,(a) provides an overview of the scanned area. It shows a small line of edge artifacts at the bottom, originating from the revolution-wise data processing. (b) shows an enlarged excerpt of (a) where the surface profile becomes visible. It is a random, but coherent profile spanning the entire FOV. We find an effective roughness of \SI{590(24)}{\pico\meter}, agreeing with the manufacturer's specification of $<\SI{800}{\pico\meter}$. We note that a strong agreement of these two values is not expected as the presented method measures the resolution limited effective roughness and not the atomic surface roughness.

We additionally observe localized features of strong amplitude sparsely scattered over the sample, as shown in (c) and (d). All features show an oscillation between a negative and positive surface level. This behavior possibly originates from the spectral filtering, as it shows a correlation to the filter's threshold frequency and only appears along the angular axis.

\subsection{Optical loss map of substrate to coating interface}
\begin{figure}[htb]
	\centering
	\includegraphics[width=\linewidth]{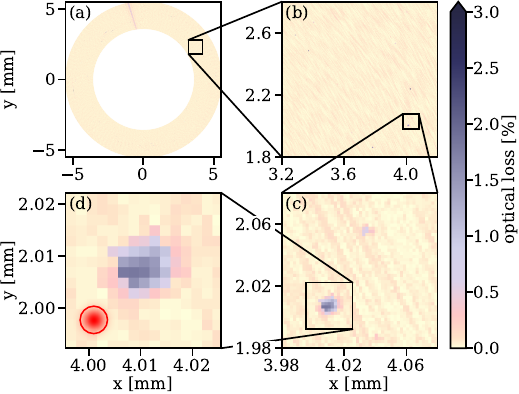}
	\caption{\textbf{Reconstructed map of measured optical loss.} Optical loss was measured per beam area $\approx\SI{22}{\micro\meter\squared}$. Lateral resolution is \SI{5.29}{\micro\meter}. (a) Overview of full scan. (b) Section of (a), enlarged by a factor of 10, showing the same spatial excerpt as fig.\,\ref{fig:effective_roughness_map}\,(b). (c) Section of (b), enlarged by a factor of 10, showing a microscopic spot of high optical loss. (d) Section of (c), enlarged by a factor of 3, comparing the feature to the probe beam's focus diameter. Feature areal density is roughly 12 features per square millimeter}
	\label{fig:loss_map}
\end{figure}

Fig.\,\ref{fig:loss_map} shows the optical loss map analogous to fig.\,\ref{fig:effective_roughness_map}. The overview (a) reveals that most of the FOV is "empty", showing a very weak background and emphasizing the mirror's quality. The top line again originates from edge artifacts. In (b) there are also some slight artifacts in angular direction at the sensitivity limit of the measurement. We again observe sparsely scattered microscopic features, as shown in (c) and (d). They cause up to \SI{65}{\percent} of optical loss per beam area of about \SI{22}{\micro\meter\squared}.

To compare the measured optical loss with the manufacturer specification of \SI{99.95}{\percent} reflectivity, we compute the mean loss over the entire FOV, excluding edge artifacts. This yields an average reflectivity of \SI{99.88(0.02)}{\percent} and \SI{99.95(0.02)}{\percent} when also excluding the microscopic spots of high loss. The latter agrees very well with the specification. One explanation of these spots could be absorbing particles enclosed at the substrate to coating interface during the coating process. In this case, it is justified to exclude them from the calculation, as they would not contribute to the reflectivity of the actual mirror surface.

\subsection{Correlations between loss and phase maps}
\begin{figure}[htb]
	\centering
	\includegraphics[width=\linewidth]{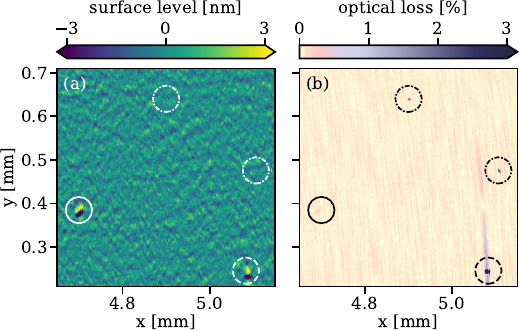}
	\caption{\textbf{Correlations between loss and phase maps.} (a) Section of the effective roughness profile. (b) Same section as in (a), showing the optical loss. The solid circle contains a feature strongly visible only in the phase map, possibly a local step in the atomic surface profile or the refractive index. The dashed circle contains a feature strongly visible in both maps, possibly a scattering defect. The dashdotted circles highlight features appearing only in the loss map, likely point absorbers.}
	\label{fig:both_maps}
\end{figure}

We find that many features appear in both maps at the same position, see fig.\,\ref{fig:both_maps}\,(a)\,-\,(c). Some features, however, only appear strongly in one map. Consequently, the observed features must fall into different categories.

Features appearing stronger in the effective roughness map are required to mostly experience a phase shift, while not causing any significant optical loss due to scattering, absorption or transmission. This would be the case for local steps in the atomic surface profile or a change in the refractive index.
If features appear equally strong in both maps, additionally, they are required to cause some optical loss for example due to optically scattering defects.
Finally, peaks predominantly visible in the optical loss map must not experience any significant phase shift. Such point-like absorbers are especially interesting for high laser power applications, as they can heat up causing local distortions of mirror surfaces and thus induce scattering losses \cite{jia2021point}.

\section{Conclusion and outlook}
We present tomographic laser scanning microscopy for profiling optical surfaces regarding phase and loss topography. We demonstrate its potential in an experimental setup, yielding partly correlated phase and optical loss maps of the same FOV and topology. These images cover an area of \SI{56}{\milli\meter\squared}, while having micrometer range resolution and close to shot-noise limited phase sensitivity.

On our sample mirror, we observe a cohesive but random reflection phase profile corresponding to an effective surface roughness. In both the phase map and the optical loss map, we find point-like features. Our technique allows for correlations between phase and loss topography to be revealed, which should provide a better understanding of limiting imperfections in optical surfaces. Such include point absorbers, which are a current problem for high laser power applications, like gravitational wave detectors\cite{jia2021point, brooks2021point}.

Due to the spiral scanning approach, the measurement duration is short and scales with the FOV's radius instead of its area. There is no conceptual limit to the FOV size and using optimized algorithms, real-time data processing in parallel to the measurement is realistic. This makes the presented method a viable option even for large optics like those used in gravitational wave detectors or industrial wafers, without compromising resolution.\\
\\
\large{\textbf{Disclosures.}} The authors declare no conflicts of interest.\\
\\
\large{\textbf{Funding.}} This work was funded by the Deutsche Forschungsgemeinschaft (DFG, German Research Foundation) under Germany’s Excellence Strategy EXC 2121 ‘Quantum Universe’ 390833306.\\
\\
\large{\textbf{Data availability.}} Data underlying the results presented in this paper are not publicly available at this time but may be obtained from the authors upon reasonable request.

\bibliographystyle{ieeetr}
{\normalsize\bibliography{bib.bib}}

\end{document}